\documentclass[times, twoside, watermark]{arxiv}

\leadauthor{Gkeka} 

\begin{document}

\title{Machine learning force fields and coarse-grained variables in molecular dynamics: application to materials and biological systems}
\shorttitle{ML in MD}

\author[1,\Letter]{Paraskevi Gkeka}
\author[2,3,\Letter]{Gabriel Stoltz}
\author[4]{Amir Barati Farimani}
\author[1]{Zineb Belkacemi}
\author[5]{Michele Ceriotti}
\author[6]{John Chodera}
\author[7]{Aaron R.\ Dinner}
\author[8]{Andrew Ferguson}
\author[9]{Jean-Bernard Maillet}
\author[10]{Herv\'e Minoux}
\author[11]{Christine Peter}
\author[12]{Fabio Pietrucci}
\author[6]{Ana Silveira}
\author[13]{Alexandre Tkatchenko}
\author[14]{Zofia Trstanova}
\author[6]{Rafal Wiewiora}
\author[2,3,\Letter]{Tony Leli\`evre}

\affil[1]{Structure Design and Informatics, Sanofi R\&D, 91385 Chilly-Mazarin, France}
\affil[2]{Ecole des Ponts ParisTech, France}
\affil[3]{Matherials project-team, Inria Paris, France}
\affil[4]{Carnegie Mellon University, USA}
\affil[5]{Laboratory of Computational Science and Modelling, Institute of Materials, \'Ecole Polytechnique F\'ed\'erale de Lausanne, Switzerland}
\affil[6]{Sloan Kettering Institute, USA}
\affil[7]{Department of Chemistry, The University of Chicago, Chicago, Illinois 60637, USA}
\affil[8]{Pritzker School of Molecular Engineering, 5640 South Ellis Avenue, University of Chicago, Chicago, Illinois 60637, USA}
\affil[9]{CEA-DAM, DIF, France}
\affil[10]{Structure Design and Informatics, Sanofi R\&D, 94403 Vitry-sur-Seine, France}
\affil[11]{University of Konstanz, Germany}
\affil[12]{Sorbonne Université, UMR CNRS 7590, MNHN, Institut de Minéralogie, de Physique des Matériaux et de Cosmochimie, 75005 Paris, France}
\affil[13]{Physics and Materials Science Research Unit, University of Luxembourg, Luxembourg}
\affil[14]{School of Mathematics, The University of Edinburgh, UK}

\maketitle

\begin{abstract}
  Machine learning encompasses a set of tools and algorithms which are now becoming popular in almost all scientific and technological fields. This is true for molecular dynamics as well, where machine learning offers promises of extracting valuable information from the enormous amounts of data generated by simulation of complex systems. We provide here a review of our current understanding of goals, benefits, and limitations of machine learning techniques for computational studies on atomistic systems, focusing on the construction of empirical force fields from ab-initio databases and the determination of reaction coordinates for free energy computation and enhanced sampling.
\end {abstract}


\begin{corrauthor}
Paraskevi.Gkeka\at sanofi.com, gabriel.stoltz\at enpc.fr, tony.lelievre\at enpc.fr
\end{corrauthor}

\section{Introduction}
\label{S:Intro}

\noindent The atomistic representation of physical systems offers a precise description of matter. Simplified models based on coarse-grained (CG) representations offer an alternative that can significantly aid in the understanding of the physical properties of the systems under consideration. Such representations can also be used as a surrogate model for enhanced sampling methods (e.g. sampling large conformational changes using reduced models).

Both in the case of biochemical systems as well as in materials, a CG description can be based on distance metrics for structural clustering~\cite{Zhang2019JCP}, as well as on reaction coordinates: for instance, the conformational changes of a complex molecule can be modeled by a few key functions of the atomic positions, while a phase transition can be described by a change of the average atomic coordination or box shape. In condensed matter physics, atomic descriptors are employed to summarize the key features of atomic configurations in order to predict forces and energies~\cite{Behler2011JCP,Bartok2013PRB}. 

In the past, reaction coordinates were defined using empirical methods and chemical intuition, while more systematic approaches were employed for the definition of atomic descriptors \cite{Wales2015, Baron2016}. During the last decade, the return and rise of Machine Learning (ML) techniques have initiated many efforts focusing on automating the definition of reaction coordinates or descriptors that are able to successfully describe the underlying atomic systems \cite{McGibbon2017,Chmiela-NatComm,Wang2019,Hase2019}. The employed methods, both supervised and unsupervised, vary. The most commonly used methods for the identification of reaction coordinates include Principal Component Analysis (PCA) \cite{jolliffe2002principal}, diffusion maps \cite{Coifman2006,coifman2008diffusion}, and auto-encoders \cite{chen2018molecular,chen2018collective,ribeiro2018reweighted,Wehmeyer2018}. For atomic descriptors, common choices are based on a judicious use of adjacency matrices and their generalizations, or on a large set of feature vectors based on a set of basis functions. 

We are witnessing many current attempts for automatically devising intuition-free collective variables, in particular for drug discovery applications \cite{ma2005automatic,chen2018molecular}. Although the initially very high hopes raised by numerical potentials are now mitigated, there have been quite a few systematic studies on the quality of the descriptors obtained by these approaches \cite{Shapeev2016,CLO18}.

A recent CECAM (Center Europ\'een de Calcul Atomique et Mol\'eculaire) discussion meeting\footnote{\noindent See the conference website:\newline {\tt https://cermics-lab.enpc.fr/cecam\_ml\_md/}} brought together a diverse audience of 29 participants from various scientific fields, including chemistry, drug design, condensed matter physics, materials science, and mathematics, to exchange about state-of-the-art techniques for automatically building coarse-grained information on molecular systems. In particular, we believe that the viewpoint and experience of condensed matter physicists in devising atomic descriptors could prove useful insights in devising reaction coordinates in a more systematic way. Mathematics offer, in this framework, a common language for the discussion. One distinctive feature of this CECAM meeting is that the emphasis was on the technical details of the underlying numerical methods.

In the current review, we discuss the following highlights of the meeting:
\begin{itemize}
\item \textbf{Machine learning force fields and Potential of Mean Force.}
ML techniques have been recently employed in the development of force field (FF) parameters based on quantum-mechanical calculations. More generally, ML techniques can be used to define a surrogate model of any quantity that could be obtained from a quantum chemical calculation, as a function of atomic coordinates (e.g. NMR chemical shieldings, IR dipole moments, ...), making it possible to obtain an accurate estimate of experimental observables. Such models are beginning to find merit due to their accuracy and versatility. In Section~\ref{sec:ML_FF}, we review the factors that play an important role in the accuracy and transferability of a force field. Specifically, we report the importance of the input database and the choice of the regression method for the force field construction. The use of prior physico-chemical knowledge in this construction of ML potentials is also discussed.

\item \textbf{Dimensionality reduction and identification of meaningful collective variables.} 
Another important issue discussed during the CECAM meeting is the dimensionality reduction and the identification of meaningful CVs using ML techniques (see Section~\ref{S:ML-CVs}). We considered the case when  this identification relies on a database which covers the full configuration space of the system under study (obtained for instance by high temperature sampling, steered molecular dynamics, etc), and the case when the data is restricted to a metastable state. Once a reaction coordinate is found, the question of devising a good effective model along this coordinate can also be addressed using machine learning techniques: either approximate free energies (for example by potentials involving only 2, 3 or 4 body interactions), or approximate the terms in the effective dynamics, namely the drift, diffusion coefficient, metric tensor and memory terms, for example using projections {\em \`a la} Mori-Zwanzig.

\item \textbf{Applications of machine learning techniques in biological systems and drug discovery.}
In Section~\ref{S:Applications}, we discuss some ``real world" applications, where MD simulations coupled with ML techniques enable us to understand the biological complexity at the atomic and molecular levels and provide us with interesting insights about the thermodynamic and mechanistic behaviour of biological processes. In particular, we highlight some examples of ML approaches applied in clustering and construction of Markov state models, we describe how ML methods facilitate enhanced sampling protocols through the use of efficient CVs and we mention some possible applications in the drug discovery process. These examples illustrate the current state and potential of the field of ML in the study of biological systems and drug discovery.

\end{itemize}
We close the review with some perspectives in Section~\ref{S:Perspective}.

\section{Machine learning force fields and Potential of Mean Force}
\label{sec:ML_FF}

\noindent Interactions between atoms are often modeled using empirical potentials with some prescribed functional forms, as suggested by physical considerations. This provides computationally cheap (with a cost scaling linearly with the number of atoms) but somewhat inaccurate potentials. On the contrary, ab-initio approaches provide more reliable, less uncertain force fields, at the expense however of a large computational cost (typically scaling as the number of electrons to the power~3). The promise of machine learning for force field computations is to predict forces and energies with accuracy arbitrary close to the level of ab-initio approaches~\cite{Lunghu2019SciAdv}, but with a much smaller computational cost and scaling as a function of the number of atoms. Ideally, these force fields should be able to describe chemical reactions. This is typically done in practice by setting up a database of configurations with associated forces and energies, summarizing atomic configurations through some descriptors of the local environment, and predicting the forces and energies from these descriptors through a function which has been trained by some (nonlinear) regression procedure to provide good results on the database. The resulting potential is called a ``numerical potential".

There are three different factors to discuss the success of ML methods, whose relative importance depend on the aims of the user: accuracy, computational cost, and transferability. The latter concept means that a numerical potential computed for a given material in a given thermodynamic range, can be used outside the fitting domain -- for instance because it is used for other materials and systems than the ones it was trained on, and/or in a different thermodynamic range than the one considered for the configurations in the database.

We first discuss in this section elements on the choice of the database, see Section~\ref{sec:database}. We next present various choices for the descriptors and for associated ML regression methods, see Section~\ref{sec:regression}. We then discuss in Section~\ref{sec:synergy_physics} how to incorporate physical insights in order to improve ML techniques, and we give some perspectives in Section~\ref{sec:FF_perspectives}. We end the section by mentioning how ML approaches can also be used to derive CG potentials, see Section~\ref{sec:CG_FF}: in this perspective, empirical force fields for all atom models are seen as the reference (they are the counterpart of ab-initio databases in this context), and effective force fields describing the interaction of coarse-grained variables are sought.

\subsection{Setting up a database} 
\label{sec:database}

\noindent One of the key factors that affects the accuracy and transferability of a force field is the database used for its construction. This database defines the envelope of confidence (applicability domain) for the potential as the subsequent regression method is efficient in interpolation. It is often the case that a numerical potential has a poor transferability. Therefore, for condensed matter systems, the database should sample the region of interest, i.e., the thermodynamic conditions where the potential is going to be used. However, this representative part of the configurational space covers only a small fraction of the overall available space. Hence, a systematic exploration is impossible, and physical intuition is often used to constrain the search of new interesting configurations for learning. This makes the construction of the database a rather laborious process. A first application of `active learning' in this process, also still hand made, is proposed by~\citeauthor{Artrith2012PRB} in Ref. \citenum{Artrith2012PRB}: two different neural networks are optimized on the same database and, in case their predictions on a new configuration differ too much this configuration should be included in the database. Active learning, based on outlier detection (i.e., definition of a metric to detect parameters corresponding to some extrapolation) is now routinely employed during the database construction~\cite{Podryabinkin2019PRB}. In this way, force field accuracy can be improved during the training procedure~\cite{GUBAEV2019148} and the domain of applicability could be extended~\cite{Huan2019JPhysChemC}. The bottom line is that `on the fly' learning~\cite{Jinnouchi2019PRB} enables to perform optimization and prediction at the same time~\cite{Deringer2018Faraday}. Typically, a trade-off has to be found between the transferability of a potential (its robustness to changes in the database) and its accuracy.

The representation of the database should also be meaningful: finding a proper space for this representation allows to define an envelope of confidence for the potential. When the potential is used, each new configuration can rapidly be plotted in this space to check if it belongs to the database envelope (applicability domain), i.e., if the potential is used in interpolation or in extrapolation. It then becomes a useful criterion for outlier detection.

What is globally accepted is that the methods 
should systematically be validated on test data, different from the training data. In any case, one should be very careful about the quality of the model for extrapolation.

\subsection{Descriptors and regression methods}
\label{sec:regression}

\noindent We present in this section the technical approaches to fit a potential on a database. We distinguish the representation of the atomic configurations through descriptors, and the subsequent regression allowing to fit the parameters of the chosen model. Typically, a very simple descriptor, based on physical/chemical intuition or moment estimates for atomic densities, should be combined with a complex regression such as a neural network; on the other hand, more educated descriptors, for instance based on convolutional neural networks and a scattering transform~\cite{EEHMT18}, can be fed into quite simple (bi)linear regression models.

\subsubsection{Representing atomic configurations}
\label{sec:descriptors}

It is almost never appropriate to use the Cartesian coordinates of atoms in a structure as the input of a machine-learning scheme~\cite{Ferre2017JCP}, because Cartesian coordinates do not conform with the invariance of the target properties, e.g. permutation of the indices of identical atoms, rigid translations, rotations and reflections. For this reason, several different schemes have been devised to map atomic configurations onto vectors of features that fulfil these symmetry requirements. Usually, it is desirable for this mapping to be differentiable and smooth, particularly in applications where one needs to compute forces as the derivative of a machine-learning potential or CG force field.  

One can roughly partition methods to represent atomic configurations into two classes. \emph{Descriptors} are often highly simplified representations of a structure, usually of much smaller dimensionality than the number of degrees of freedom and incorporating some degree of chemical intuition, or a heuristic understanding of the behavior of the system being studied. Cheminformatics schemes to characterise the connectivity of a molecule, such as SMILES~\cite{Weininger1988} strings, are useful when dealing with databases of organic compounds. Steinhardt parameters~\cite{Steinhardt1983} are often used to characterize the coordination of liquids and solids. Backbone dihedral angles, or more complex indicators of secondary structure~\cite{Pietrucci2009} can be utilized to discard information on the side chains of polypeptides. The dimensionality reduction that is intrinsic to this family of methods typically induce loss of information, which may be desirable (when it discards irrelevant details) or problematic: in the latter case, it is often more effective to use a more complete description and then proceed with an automatic dimensionality reduction algorithm, some of which will be discussed in Section~\ref{S:ML-CVs}. 

\emph{Representations}, on the other hand, attempt to provide a complete description of a configuration. This family of features is typically used when building regression models for energy and properties. Most of the time (particularly for condensed-phase applications, but often also for isolated molecules) representations are not built for an entire structure, but are instead used to describe atom-centered environments. This is advantageous, because - by representing a structure as a collection of compact groups of atoms, and assuming that the overall property can be computed as a sum of local contributions - it becomes possible to train models that can be easily transferred between systems of different sizes, and from simple to more complex configurations. Many of these systematic representations - including e.g., SOAP (bi)spectrum~\cite{GAP}, Behler-Parrinello symmetry functions~\cite{Behler-Parrinello}, moment tensor potentials~\cite{Shapeev2016}, FCHL kernels~\cite{Faber2018} - can be seen as projections on different basis of n-body correlation functions~\cite{Willatt2019}, and offer a systematic and completely general way to describe atomic configurations, that can be applied equally well to condensed phases, gas-phase molecules and polypeptides~\cite{Bartok2017}.

\subsubsection{Choosing the regression method}

\noindent Once the atomic descriptor has been chosen, the choice of the regression method to determine the force field is crucial and greatly depends on the system under study~\cite{Zuo2020JPCA}. A distinction should be made between learning based on neural networks, and other regression methods based on kernels or (bi)linear methods. Training neural networks is a complex non-convex optimization problem in very high dimension (generally thousands of parameters are needed to parameterize the networks under consideration). Already the computation of the gradient of the objective function is non trivial and relies on clever numerical tricks, such as backpropagation. Kernel-based methods or (bi)linear regression techniques lead, on the other hand, to much better behaved optimization problems, which can even be solved analytically through some matrix inversion on the Euler equation defining the minimizer. 

The choice of the regression method also determines whether error estimators are available. For example a variance can be associated with a prediction when a kernel method is used, whereas error quantification is harder using neural networks. Moreover, the robustness of the potential depends on the regression method and its associated regularization (used to alleviate overfitting issues). A simple (bi)linear method may be less accurate but more robust. It may also be sufficient if the descriptors already provide enough information on the system, as is the case for the descriptors obtained via convolutional neural networks in Ref.~\citenum{EEHMT18}.

In principle, both neural network (NN) and non-linear kernel regression models are sufficiently sophisticated to obtain a trustworthy representation of scalar potential-energy surfaces (PES) or vector force fields of arbitrary complexity. However, in practice, choices have to be made for the similarity measure between atomic configurations (in both kernel regression methods and NN) or for the architecture of the neural network. The optimal choices are not the same for different systems, i.e., descriptors/parameters that work well for solids are not easily transferable to biological molecules and vice versa. Hence, many ML developments are currently specific to either organic molecules or materials. That being said, there is currently a growing interest in understanding the advantages and limitations of the different existing approaches~\cite{Shapeev2016,EEHMT18,Behler-Parrinello,GAP,Behler-review,Bowman-review,Chmiela-SciAdv,DTNN} and developing truly general frameworks for learning complex PES or force fields that work seamlessly for both organic and inorganic matter.

\subsubsection{Current methods and their performances}

\noindent We list some key methods in Table~\ref{tab:regression_ff}. The first successful ML approaches were developed to describe PES of defectless materials and their surfaces~\cite{Behler-Parrinello,GAP,Behler-review} with the goal to enable efficient and accurate Molecular dynamics (MD) of large supercells of elementary or binary materials. The Behler-Parrinello NN approach~\cite{Behler-Parrinello} or the kernel-based GAP approach of Csanyi~\cite{GAP} are both able to achieve accuracies of 1-2 meV/atom for some solids (C, Si, Cu, TiO2, among others). There are several key differences between these two methods, the main ones being the NN vs kernel approach and the different similarity measures between atomic configurations. Both approaches typically require on the order of tens to hundreds of thousands reference calculations at the DFT level  for constructing the training dataset, in order to  achieve 1-2 meV/atom accuracy. Recently, PES-fitting methods based on deep networks have also been developed~\cite{DTNN,SchNet}. These approaches often do not require any \textit{a priori} definition of the similarity measure; they are instead able to learn the similarity measure from the training data.

\renewcommand{\arraystretch}{1.5}

\begin{table*}[tb] \centering
    \begin{tabular}{| l l c|}
        \hline
       \textbf{Method} & \textbf{Short description} & \textbf{Ref.} \\
       \hline 
       Kernel-based Gaussian approximation & Combines a structural descriptor and a kernel establishing & \citenum{GAP} \\
         potentials (GAP)                  & the link between structure and energy &   \\
       \hline
       Behler-Parrinello NN & Feed-forward NNs for each atom. The potential energy is &  \citenum{Behler-Parrinello,Behler-review} \\
       &  constructed as the sum of local atomic energies &  \\
        \hline
       Deep NN (DTNN) & No a priori similarity definition needed, similarity is learned & \citenum{DTNN,SchNet}\\
        \hline
       Permutationally-invariant polynomials (PIP) & Uses polynomials of Morse variables in fitting PES & \citenum{Bowman-review, Chen2019} \\
      \hline 
       Gradient-domain ML (GDML) & Learns an explicit FF and obtains the PES via integration & \citenum{Chmiela-SciAdv,Chmiela-NatComm}\\
       \hline
    \end{tabular}
    \caption{Summary of some key learning methods for force field (FF) development.}
    \label{tab:regression_ff}
\end{table*}

Constructing ML models for organic molecules is a field that faces somewhat different challenges compared to ML models for solids and materials. While DFT calculations are often deemed to provide sufficiently accurate reference data for solids, this is not the case for organic molecules. The ``gold standard'' is coupled cluster CCSD(T) computations. Quantum-chemical CCSD(T) calculations are however computationally expensive and it is only possible to carry hundreds of such calculations even for simple  molecules such as aspirin. Early successful non-linear PES models were based on permutationally-invariant polynomials (PIP)~\cite{Bowman-review}. More recent developments include the so-called gradient-domain machine learning (GDML) approach~\cite{Chmiela-SciAdv,Chmiela-NatComm} for constructing molecular force fields. The GDML approach learns an explicit force field and obtains the PES via integration, instead of the more conventional approach to learning a PES and then taking its gradient to drive MD. This has two advantages: (i) the usage of an explicit Hessian kernel that provides the maximum flexibility, minimizes noise and prevents artifacts between forces and energies in the learning process; (ii) a significant gain in data efficiency, since globally accurate force fields for small molecules (accuracy of 0.2 kcal/mol and 1 kcal/mol/{\AA}) can now be constructed using only a few hundred molecular conformations for training. This data efficiency currently enables the construction of essentially exact force fields for molecules with up to 30-40 atoms~\cite{Chmiela-NatComm}.

\subsection{Synergy between physics, chemistry, mathematics and ML approaches}
\label{sec:synergy_physics}
\noindent ML approaches used to construct accurate PES and force fields have already been successful and have enabled simulations of molecules and materials that were previously considered impossible. Ultimately, it would be worthwhile to achieve an optimal balance between physics-based models and ML approaches to enable not only faster and more accurate simulations, but also obtain insights into interactions of complex quantum-mechanical molecules and materials. For example, the GAP, Behler-Parrinello, GDML, and PIP approaches discussed above already incorporate translational, rotational, and permutational symmetries of molecules and materials in their internal representation of atomic interactions. Such symmetries were also made precise in the mathematical literature~\cite{Shapeev2016}. In addition, by learning simultaneously energy and forces such that the latter are (minus) the gradient of the former, all of these methods enforce exactly energy conservation.

However, many more physical symmetries can and should be incorporated in ML approaches. For example, exact constraints are known for asymptotic forms of atomic interaction potentials. Also, some analytic and empirical results are known for series expansions of interatomic potentials. Finally, there are mathematical results which provide rigorous statements on the behavior of the potential energy functions in terms of the locality of the interactions~\cite{CLO18}. The incorporation of such prior knowledge could improve the efficiency and accuracy of ML potentials and ultimately also lead to novel analysis tools that offer new insights into the complex nature of atomic interactions~\cite{Deringer2017PRB}. 

It is also worth noting that electronic interactions in complex molecules and materials can be rather long-ranged. For example, electrostatic interactions and plasmon-like electronic fluctuations in molecules and nanostructures can lead to interatomic potentials extending to at least 20-30 nanometers~\cite{Science-MBD,ChemRev-vdW2017}. Most current ML models explicitly or implicitly cut off interactions at an interatomic distance of 5-6 {\AA}. Hence, by construction, these ML approaches are not able to capture interactions extending over larger length scales. For this reason, it is ultimately necessary to couple ML approaches that excel at capturing complex short-range chemical bonding with explicit physics-based approaches to non-covalent interactions. It is important to note that such physics-based models can also employ ML approaches to learn short-range interaction parameters based on datasets of electrostatic moments and polarizabilities. The recently developed IPML approach lies the foundation for unifying ML force fields and physics-based interatomic potentials~\cite{IPML}. An alternative approach based on the definition of structure representations that incorporate long-range correlations with the correct asymptotic behavior~\cite{Grisafi2019} can simplify the simultaneous description of the multiple length scales contributing to molecular interactions.

\subsection{Perspectives for ML approaches to the determination of force fields}
\label{sec:FF_perspectives}

\noindent We gather in this section some mathematical and numerical perspectives, as well as open problems, on ML methods for force fields:
\begin{itemize}
\item A first perspective is the use of ML to learn the difference between already acceptable empirical force fields and DFT models, as some form of preconditioning. Such an approach greatly depends on the regression method. For example, for kernel methods, it has been shown that a potential can be built on top of pre-existing two-body and three-body classical potentials, improving the overall accuracy \cite{Glielmo2018,Veit2019}. On the contrary, fitting differences between a good classical potential and an ab-initio potential with a linear regression yields very poor results, since the difference is small (almost noisy) and rugged (not smooth). It is observed that a simpler starting guess, such as the Ziegler--Biersack--Littmark potential~\cite{ZBL}, yields better results, since this increases the numerical stability and improves the accuracy.
\item A question related to the robustness of these learning techniques is whether it would make sense to optimize potentials on a Pareto curve, where various properties of interest are weighted in different manners in the cost function. Indeed, the optimization is usually performed on a multi-objective cost function (including energy, force, stress, and sometimes bond distances, ...). The so-obtained potential is a result of the user arbitrary choice of the weighting parameters -- infinitely many `optimal' potentials can be obtained depending on the choice of the weights. The naturally rising question here is: is it possible to have a unified way of defining cost functions? 
\item  An important practical concern is the sensitivity of the learnt parameters relatively upon the data (for instance depending on the fraction of elements used for training vs. testing). 
\item  Another more theoretical question is: What is the numerical stability induced by machine learning potentials on the time integration of Hamiltonian dynamics and its variations? Indeed, some preliminary results suggest that machine learning potentials may be smoother than current empirical potentials.
\item For reasons which remain to elucidate, predicting intensive (as opposed to extensive) properties seems to be very challenging.
\end{itemize}

\subsection{Bottom-up coarse-graining force fields: From PES to FES}
\label{sec:CG_FF}

\begin{figure}[t]
	\includegraphics[width=\linewidth]{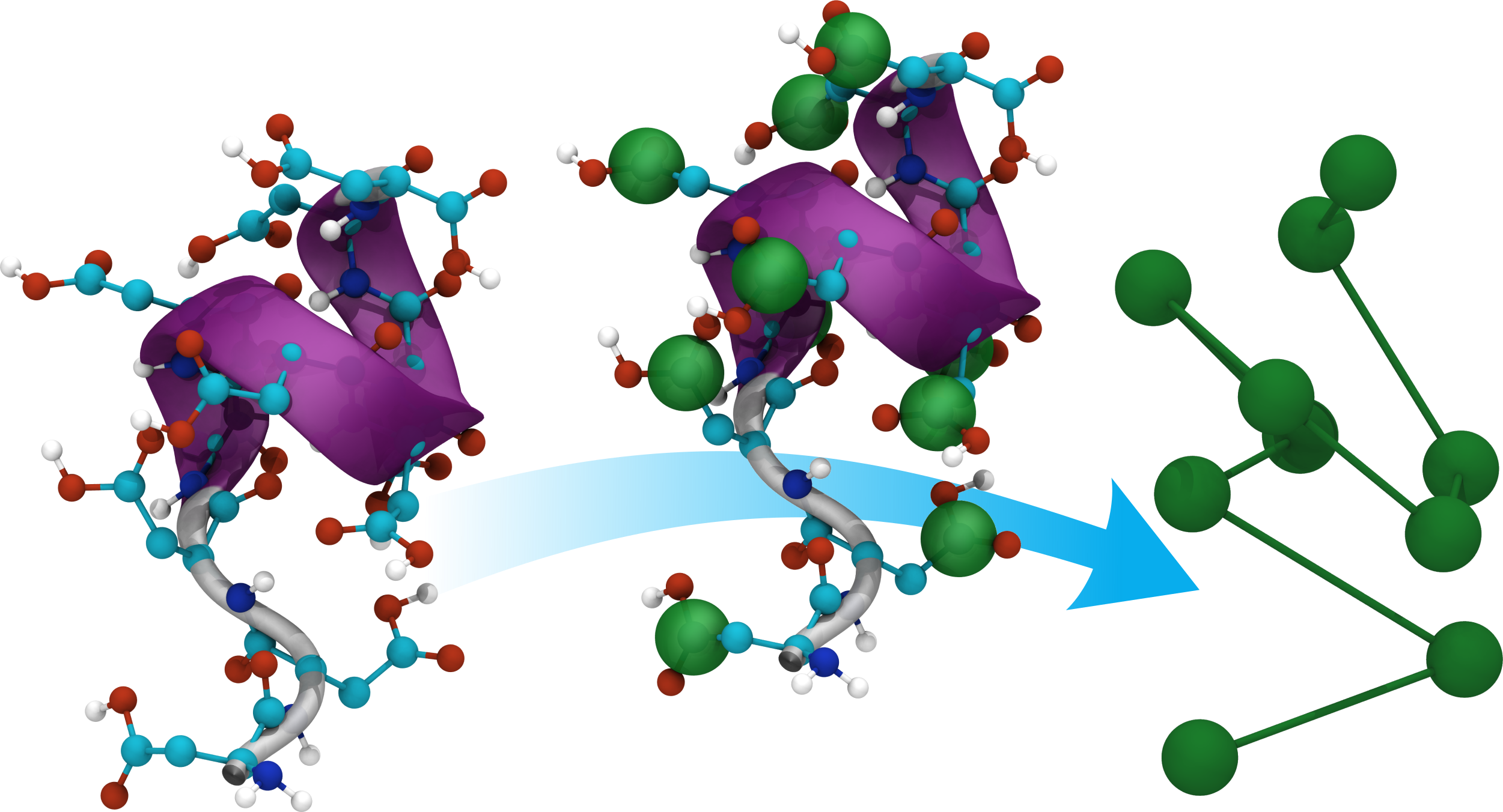}
    \centering
	\caption{Particle-based coarse-graining: high dimensional free energy surfaces (FES) can be extract from atomistic data and used as a basis for CG models \cite{Lemke2017_NN-CG, Hunkler2019}.} 
	\label{fig:CG}
\end{figure}

\noindent A classical particle-based coarse grained (CG) simulation model, where several atoms are grouped together, can be viewed as a reduction of the dimensionality of the classical phase space (see Figure \ref{fig:CG}). It requires the determination of an effective Hamiltonian that allows the model to explore the phase space in the same way as an atomistic simulation would. Thus, in the so-called bottom up coarse-graining strategies, the interactions in the CG model are devised such that an accurate representation of a (known) atomistic sampling of the configurational phase space (mapped to the CG representation) is achieved. These methods use the underlying multidimensional potential of mean force (PMF) derived from the atomistic simulation data as parameterization target, i.e., they try to reproduce a (typically high-dimensional) free-energy surface (FES) as opposed to a PES. Naturally, this is of particular relevance to the simulation of soft matter problems such as liquid state systems, soft materials and biological systems, where entropic effects, disorder and heterogeneity dominate the overall properties of the system.

Free energies and potentials of mean force are not a direct output of a MD simulation. They can be calculated by Boltzmann inversion of a (high-dimensional) probability density distribution obtained from sampling configurations in phase space or from mean forces acting on the interaction sites in the CG representation. In the past, several bottom-up coarse-graining methods have been derived which - while all aiming for an effective Hamiltonian that approximates a multidimensional PMF/FES - differ in terms of both the actual parameterization target (multidimensional PMFs/probability density distributions, structure functions as low-dimensional representations of these PMFs; mean forces in the direction of selected CVs or relative entropies) and the type of CG interactions which are  typically represented by low-dimensional potentials, i.e., pair interactions, or three-body interactions) \cite{Peter2009,Rudzinski2011,Noid2013_review,Potestio2014_review,Shell2015_review}. Since these coarse-graining methods derive interactions from atomistic reference simulations, they are intrinsically data driven. Consequently, ML-based approaches yield new types of reference atomistic data and new types of CG interactions and parameterization methods. On the one hand, ML methods can be used to determine dimensionality-reduced representations of the phase space and to derive or validate CG models by matching the sampling of a (relatively complex) FES as opposed to low-dimensional target functions/properties. On the other hand, ML methods can also be employed to identify suitable CVs that describe the states and the dynamics of a system, which can then either be directly used in the CG potentials or be employed to identify optimal CG representations and learn CG interactions. This is discussed at length in Section~\ref{S:ML-CVs}.

Following the methodology of inferring all-atom potential energy functions from corresponding quantum mechanical data, John and Csanyi have extended the Gaussian Approximation Potential (GAP-CG) approach to coarse-graining of simple liquid systems \cite{John2017_GAPCG}. In this case, the many-body PMF is described via local multibody terms, based on local descriptors and multidimensional functions which are determined by Gaussian process regression from atomistic training data (instantaneous collective forces or mean forces).
In a similar vein, Zhang et al. developed a scheme, called the Deep Coarse-Grained Potential (DeePCG), which uses  a NN to construct a many-body CG potential for liquid water \cite{Zhang2018_DEEPCG}. The network is trained with atomistic data in a manner similar to the force matching in the multi-scale coarse-graining method \cite{Noid2008}, and in such a way that it preserves the natural symmetries of the system. 
While the described two methods are related to the force-matching type of bottom-up coarse-graining and use ML to significantly extend the complexity of the CG interactions, Lemke and Peter follow a different strategy \cite{Lemke2017_NN-CG}. 
A NN is used to extract high-dimensional FES from atomistic MD simulation trajectories. The NN is trained to predict conformational free energies by creating a classification problem between real MD conformations and fake conformations of a known distribution. 
With such a classification based procedure it is possible to train the NN to return probability densities without requiring any binning or normalization -- which circumvents the problem of binning in high dimensional space~\cite{Garrido_Classification}. By using the NN probability densities directly in a Monte Carlo type of sampling of conformations, a (relatively) high-dimensional FES is thus used as effective CG Hamiltonian. This NN network model was successfully tested for several homo-oligopeptides \cite{Hunkler2019}.
By employing a convolutional NN architecture, the NN model could be simultaneously trained on data of different chain lengths and could even  make meaningful predictions for polymers with chain lengths different from the ones in the training data. Thus, such an approach is promising for the simulation of polymer systems where naturally training data are restricted to chain lengths that are  shorter than the intended polymers.

Coarse-graining of potential energy functions into free energy type interactions has a well founded statistical interpretation. A difficult question is however whether some dynamical properties are also preserved in this coarse-graining process, and to which extent.

\section{Dimensionality reduction and identification of collective variables}
\label{S:ML-CVs}

\noindent The objective of this section is to discuss various techniques to identify collective variables. After some general considerations in Section~\ref{sec:31}, we first present the main two ideas to build collective variables in Section~\ref{sec:32}, namely looking for high-variance or slow degrees of freedom. We then discuss how this can be used to enhance the sampling of the canonical ensemble on the example of diffusion maps in Section~\ref{sec:33}, before discussing dynamical aspects in Sections~\ref{sec:34} and~\ref{sec:35}.

\subsection{General considerations}
\label{sec:31}
\noindent Molecular systems are characterized by the fact that their long-time dynamical behavior is typically governed by a small number of emergent collective variables (CVs) \cite{ferguson2010systematic,ferguson2011nonlinear,wang2018nonlinear}. These collective modes arise from cooperative couplings between the constituent atoms induced by interatomic forces (e.g., covalent bonds, electrostatics, van der Waals interactions) and possibly external fields (e.g., electric fields, hydrodynamic flows), and which render the effective dimensionality of the system far lower than that of the full-dimensional phase space in which the system Hamiltonian and equations of motion are formulated \cite{ferguson2011nonlinear,wang2018nonlinear}. In a dynamical systems sense, the long-time evolution of the system is restrained to a low-dimensional attractor or intrinsic manifold and its dynamics over these time scales may be described within the Mori-Zwanzig projection operator formalism as evolving within a subspace of slow collective variables to which the remaining degrees of freedom are effectively slaved \cite{ferguson2011nonlinear}. 

Traditional unbiased MD is not able to efficiently explore the whole kinetic landscape with time scales spanning over orders of magnitude, from picoseconds to milliseconds. In this scenario, one relies on extensive simulations together with some clever strategy to escape metastable states. Such a strategy can only be devised if one is able to identify what defines a ``long-lived'' state, which is equivalent to discovering meaningful collective variables (CVs) or reaction coordinates \cite{Pietrucci_p2017strategies}.

The methods described below aim at finding these CVs or states. As will become clear later, depending on the objective, the focus may be different: gain insight/intuition on the system, bias to exit metastable states, compute a free energy profile, set up a coarse-grained dynamics simulation, cluster/classify configurations, etc.

\subsection{Data-driven discovery of high-variance and slow collective variables}
\label{sec:32}
\noindent The inherently multi-body and emergent nature of the CVs means that they are exceedingly challenging to intuit for all but the most trivial systems, and data-driven techniques present a powerful means to systematically estimate them from molecular simulation data. The origins of this data-driven approach can be traced back to pioneering work in the early 1990's by Toshiko Ichiye and Martin Karplus \cite{ichiye1991collective}, Angel Garcia \cite{garcia1992large} and Andrea Amadei, Antonius Linssen and Herman Berendsen \cite{amadei1993essential} who applied PCA to molecular simulations of protein folding. Since that time there has been an explosion of interest in the use of data science and machine learning techniques to estimate CVs from molecular simulation data and the subsequent use of these CVs to inform new understanding, perform molecular design, and guide enhanced sampling.

Data-driven CV discovery typically employs unsupervised learning techniques that seek low-dimensional parameterizations of the geometry of the data in the high-dimensional phase space of atomic coordinates \cite{ferguson2017machine}. This procedure can usually be cast as an optimization problem that maximizes some objective function, or equivalently minimizes some loss function, over the data. The techniques can be categorized into linear and nonlinear methods. Linear techniques are restricted to discovering CVs that are linear combinations of the input features, whereas nonlinear techniques can discover more general nonlinear functional relations. The more powerful and general nonlinear techniques are typically better suited to the estimation of the complex emergent CVs in molecular systems, but linear techniques should not be discounted since they are typically more robust, interpretable, and less data hungry, and can also admit nonlinearities through feature engineering or the kernel trick \cite{scholkopf2001kernel}. The importance of the choice of features in which the molecular system is represented to the CV discovery tool should not be underestimated. Feature sets that contain and foreground the important molecular behaviors and respect fundamental symmetries (e.g., translation, rotation, permutation) can be critical to the success of CV discovery (particularly in the case of linear techniques), whereas poor choices that mask or discard essential information or contain spurious symmetries can easily produce poor performance. What constitutes a good choice of feature set is strongly system dependent and is typically reliant on some combination of intuition, experience, and exploratory trial-and-improvement. We refer for example to Ref.~\citenum{sittel2018} for a discussion on the importance of the choice of the representation of the data.

\begin{figure*}[tb]
	\includegraphics[width=.7\linewidth]{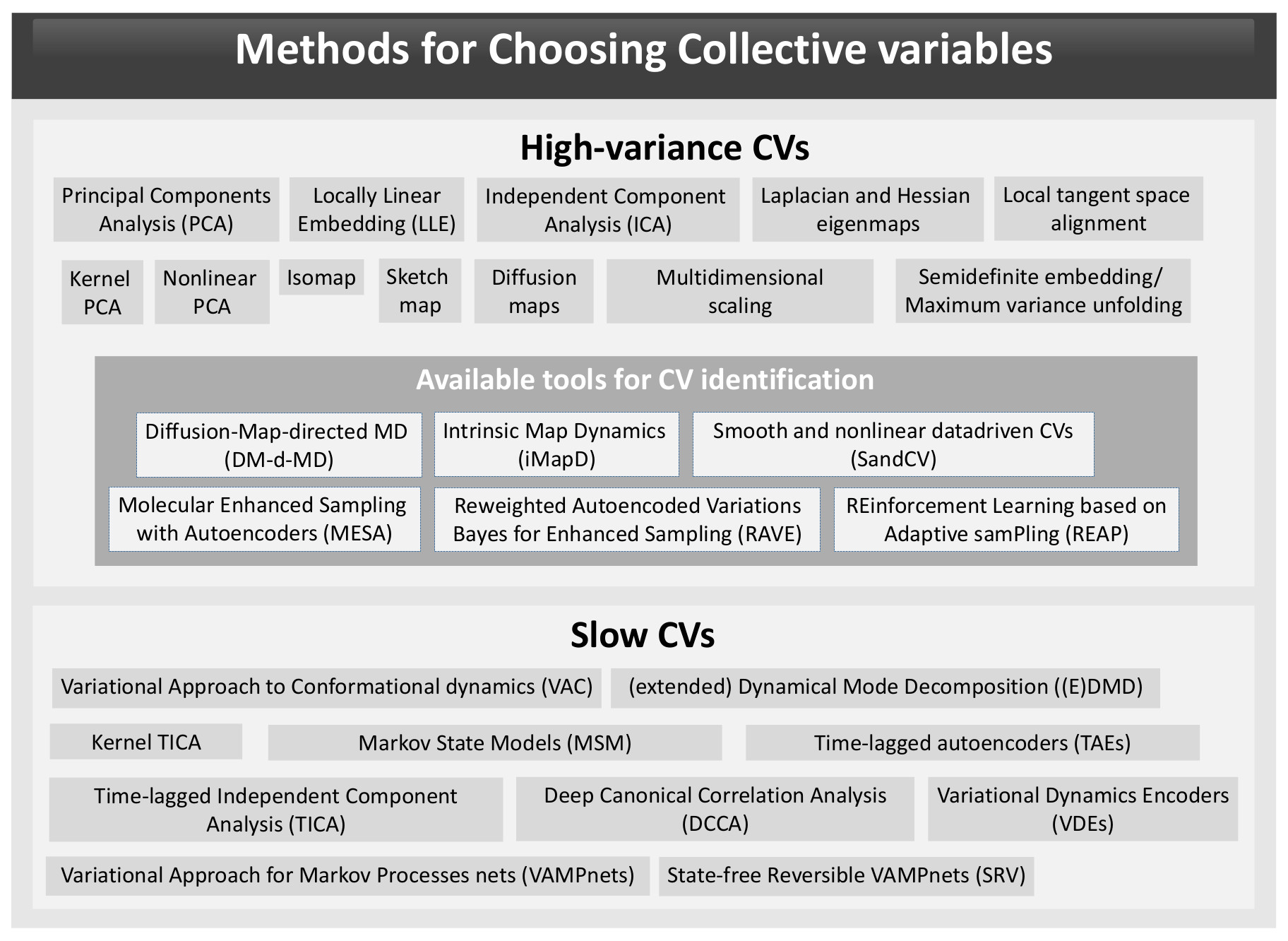}
    \centering
	\caption{Representative methods for CV identification. All related citations are in the main text.} 
	\label{fig:CV}
\end{figure*}

Although the details and specifics differ, most CV discovery techniques can be placed in one of two categories: those that seek high-variance CVs and those that seek slow CVs (see Figure~\ref{fig:CV}). 

High variance CVs maximally preserve the configurational variance in the high-dimensional data upon projection into the low-dimensional space spanned by these CVs. Slow (i.e., maximally autocorrelated) CVs define a low-dimensional space that maximally preserves the long-time kinetics of the system. Frequently the slow and high-variance collective modes are related, but this is not always the case. Importantly, the estimation of slow CVs requires data arranged in time series (e.g., MD trajectories) whereas the estimation of high-variance CVs can be applied to data sampled without temporal ordering (e.g., Monte Carlo trajectories). Notice however that methods exist to recover dynamical information according to some artificial dynamics (e.g. reversible purely diffusive dynamics) upon non-time ordered data to render it amenable to temporal analysis techniques \cite{trstanova2019local}. 
 
Let us also mention that recent advances in deep reinforcement learning (DRL) in robotics opens up new avenues for deploying DRL to atomic and molecular systems. In all DRL algorithms, a reward function, state and action space should be defined. In atomic systems, state space can be atomic coordinate, action space can be the movement of atoms, and reward can be defined as energy. DRL can be suitable replacement for finding transition paths and can potentially be used to strengthen the string or nudged-elastic-band method~\cite{jonsson1998nudged,weinan2002string}.

Before giving more details about the high-variance and slow CVs, let us mention that a widespread definition of an optimal {\em scalar-valued} reaction coordinate in the rare event-field is the committor function, i.e., in a system with two metastable states, the probability that a given atomic configuration will evolve towards the products before reaching the reactants. Such probability can in principle be estimated by generating a huge number of MD simulations from each configuration of interest: even if such a procedure cannot be applied in practice to the whole configuration space, the committor represents an ideal reaction coordinate in some sense (we refer the reader to~\cite{Weinan10} or~\cite[p.126]{lelievre2016partial} for example) and provides tests and optimization strategies for candidate CVs~ \cite{Baron2016,ma2005automatic,Weinan10,Bolhuis02,Banushkina16,jung2019artificial}.

\subsubsection{High-variance CV estimation}
\noindent The best known high-variance CV estimation technique is PCA \cite{jolliffe2002principal}, also known as the Karhunen-Lo\`eve transform \cite{loeve1955probability, sirovich1987turbulence, sirovich1987turbulence2, park1996use}, or proper orthogonal decomposition \cite{chatterjee2000introduction, liang2002proper}. This approach discovers an orthogonal transformation of the input data to define a hyperplane approximation that preserves most of the variance in the data. Popular nonlinear techniques for high-variance CV estimation include kernel and nonlinear PCA \cite{scholkopf1997kernel,kramer1991nonlinear,Nguyen2006,scholz2008nonlinear}, independent component analysis (ICA) \cite{ICA}, multidimensional scaling~\cite{borg2005modern}, sketch map~\cite{Ceriotti2011} locally linear embedding (LLE) ~\cite{roweis2000nonlinear, zhang2006mlle}, Isomap \cite{das2006low, tenenbaum2000global, Silva2002}, local tangent space alignment \cite{wang2012local}, semidefinite embedding / maximum variance unfolding \cite{weinberger2006unsupervised}, Laplacian and Hessian eigenmaps \cite{belkin2003laplacian, donoho2003hessian}, and diffusion maps ~\cite{Coifman2006,coifman2005geometric}. These approaches differ in their mathematical details, but can be broadly conceived of as nonlinear analogs of principal component analysis that pass curvilinear manifolds through the data to define nonlinear projections into a low-dimensional subspace spanned by the learned CVs. Specialized techniques for molecular simulations that integrate iterative high-variance CV discovery and accelerated sampling of configurational space have been developed in recent years \cite{chen2018molecular,chen2018collective,ribeiro2018reweighted,shamsi2017reinforcement,chiavazzo2017intrinsic,ferguson2011integrating,Tribello2012,abrams2012fly,hashemian2013modeling,li2006version,spiwok2011metadynamics,branduardi2007b,Preto2014,Zheng2013}. 

The techniques described above can be coupled with enhanced sampling methods, which use the uncovered CV's to help the system leave metastable states. In this case, one actually relies on CV estimates based on partial sampling~\cite{trstanova2019local}. Let us describe a few methods in that direction.

Diffusion-map-directed MD (DM-d-MD) uses diffusion maps to identify CVs spanning the range of explored system configurations and then initializes new simulations at the frontiers of this domain to drive sampling of new system configurations \cite{Preto2014,Zheng2013}. Intrinsic map dynamics (iMapD) employs diffusion maps to construct a nonlinear embedding of the high dimensional simulation trajectory and then uses boundary detection algorithms with a local principal components analysis to extrapolate into new regions of phase space at which to seed new simulations \cite{chiavazzo2017intrinsic}. The Smooth And Nonlinear Data-driven Collective Variables (SandCV) approach identifies nonlinear CVs using Isomap, expands them within basis functions centered on a small number of landmark points, and then passes this parameterization to the adaptive biasing force accelerated sampling technique to drive sampling along these coordinates \cite{hashemian2013modeling}. Molecular enhanced sampling with autoencoders (MESA) employs autoencoding neural networks to discover nonlinear CVs for enhanced sampling without the need for approximate basis function expansions \cite{chen2018molecular,chen2018collective}. Reweighted Autoencoded Variational Bayes for Enhanced Sampling (RAVE) employs variational autoencoders to discover nonlinear CVs that are compared at the level of their probability distributions with an ensemble of physical candidate variables to identify physical coordinates for accelerated sampling \cite{ribeiro2018reweighted}. REinforcement learning based Adaptive samPling (REAP) employs reinforcement learning to identify the dynamically-varying relative importance in driving exploration of configurational space of each CV within a candidate set and then adaptively seeds new simulations from configurations with high reward functions \cite{shamsi2017reinforcement}.

\subsubsection{Slow CV estimation}
\label{sec:slow_CV}

\noindent The identification of slow CVs is valuable and informative from many perspectives. From a mechanistic perspective, these CVs reveal the collective modes that dictate the metastable states of the system and the transitions between them. From a design perspective, they can offer a blueprint for the structural, thermodynamic, and dynamic properties of the system. From an enhanced sampling perspective, they provide good variables in which one can apply biases to accelerate barrier crossing and improve exploration of configurational phase space.

A number of approaches have been proposed to analyze MD time series to estimate slow CVs. The theoretical basis for these techniques is founded in the variational principle of conformational dynamics (VAC) \cite{noe2013variational}, or in the (extended) dynamical mode decomposition ((E)DMD) \cite{mezic2005spectral,williams2015data} that, respectively, frame the  recovery of the slow CVs as a variational optimization or regression problem \cite{Wehmeyer2018,wu2017variational}. Shortly, VAC estimates the slowest modes as linear combinations of \textit{a priori} defined basis functions of the input coordinates. In Time-lagged independent component analysis (TICA) these basis functions are the coordinates themselves \cite{noe2013variational,perez2013identification,nuske2014variational, noe2015kinetic, noe2016commute, perez2016hierarchical, Schwantes2013, klus2018data}. In Markov state models, the slow CVs are approximated in a basis of  indicator functions defined over the data \cite{wu2017variational,pande2010everything} (see also the recent special issue Ref.~\citenum{MSM_special_issue} for the latest developments on Markov state models). Perron cluster analysis can be used to reduce the large number of states uncovered by clustering methods along the trajectory, to a few metastable states, see Ref.~\citenum{schutte1999direct,deuflhard2005robust,roblitz2013fuzzy}. Combining TICA with the kernel trick yields kernel TICA (kTICA) that is capable of approximating the slow CVs with nonlinear functions of the input features \cite{schwantes2015modeling, noe2013variational}. Deep canonical correlation analysis (DCCA) \cite{andrew2013deep}, the variational approach for Markov processes nets (VAMPnets) \cite{Mardt2018}, and state-free reversible VAMPnets (SRV) \cite{chen2018srv} all employ Siamese neural networks to learn nonlinear featurizations of the input coordinates as basis functions with which to approximate the slow CVs. Time-lagged autoencoders (TAEs) employ time-delayed autoencoding neural networks to learn slow CVs into which the molecular trajectory can be projected (i.e., encoded) and also used to predict the system state at the next time increment (i.e., decoded) \cite{Wehmeyer2018}. Variational dynamics encoders (VDEs) are similar to TAEs but employ a variational as opposed to traditional autoencoding architecture that introduces stochasticity  into the decoding of the learned CVs \cite{hernandez2018variational,wayment2018note}.

Enhanced sampling can be conducted in the learned slow CVs in a similar manner to that in the high-variance CVs, but the application of artificial biasing potentials perturbs the true system dynamics and subsequent applications of slow CV estimation techniques to the biased data must compensate for this effect \cite{quer2018automatic,donati2018girsanov,donati2017girsanov}.

\subsection{Enhanced sampling using local and global diffusion maps}\label{sec:33}

\noindent Using the illustrative example of diffusions maps, we discuss in this section how to use the proposed reaction coordinate to enhance sampling and somehow perform some extrapolation procedure.
Diffusion maps are a dimensionality reduction technique which allows for identifying the slowly-evolving principal modes of high-dimensional molecular systems \cite{Coifman2006,coifman2008diffusion}. It does so by computing an approximation of a Fokker-Planck operator on the trajectory point-cloud sampled from a probability distribution (typically the Boltzmann-Gibbs distribution corresponding to prescribed temperature). The construction is based on a normalized graph Laplacian matrix. In an appropriate limit, the matrix converges to the generator of overdamped Langevin dynamics. The spectral decomposition of the diffusion map matrix thus yields an approximation of the continuous spectral problem on the point-cloud~\cite{nadler_springer} and leads to natural CVs.

Since the first appearance of diffusion maps~\cite{Coifman2006},  several improvements have been proposed including local scaling~\cite{Rohrdanz2011a}, variable bandwidth kernels~\cite{Berry2016} and target measure maps (TMDmap)~\cite{Banisch2018}. The latter scheme extends diffusion maps on point-clouds obtained from a surrogate distribution, ideally one that is easier to sample from. Based on the idea of importance sampling, it can be used on biased trajectories, and improves the accuracy and application of diffusion maps in high dimensions~\cite{Banisch2018}. 

Several algorithms have used diffusion maps to learn the CVs adaptively and thus enhance the dynamics in the learned slowest dynamics~\cite{chen2018molecular,chiavazzo2017intrinsic,Preto2014,Zheng2013}. These methods are based on iterative procedures whereby diffusion maps are employed as a tool to gradually uncover the intrinsic geometry of the local states and drive the sampling toward unexplored domains of the state space, either through sequential restarting~\cite{Zheng2013} or pushing~\cite{chiavazzo2017intrinsic} the trajectory from the border of the point-cloud in the direction given by the reduced coordinates. All these methods try to gather local information about the metastable states to drive global sampling.  
In~\cite{trstanova2019local}, the authors focused on the construction of diffusion maps within a metastable state by formalizing the concept of a local equilibrium based on the \textit{quasi-stationary distribution}~\cite{collet2012quasi}. This local equilibrium guarantees the convergence of the diffusion map within the metastable state. Moreover, the work provides the analytic form of the operator obtained when metastable trajectories are used within diffusion maps.

Finally, since the collective variables provided by diffusion maps are only defined on the sampled point cloud,  
one must apply extrapolation approaches. These might be very noisy and, more importantly, lose their meaning outside the convex hull of the point cloud. As a remedy, diffusion maps could be used as a tool to select collective variables from a database of physical reaction coordinates, similarly to~\cite{ma2005automatic}, providing more physical insight into the abstract collective variables. This approach would allow to evaluate the CV outside the point cloud and provide more physical meaning into the abstract collective variables. 

The local-global perspective has motivated a method allowing on-the-fly identification of metastable states as an ensemble of configurations along a trajectory, for which the diffusion map spectrum converges. Secondly,  an enhanced sampling algorithm based on QSD and diffusion maps has been proposed. For the latter, the main idea is a sample from the QSD allowing to build high-quality local CVs (within the metastable state) by considering the most correlated physical CVs to the diffusion coordinates. Once the best local CVs have been identified, one can use existing methods as metadynamics to enhance the sampling, effectively driving the dynamics to exit the metastable state. The authors in \cite{trstanova2019local} demonstrate this idea on a toy-model example showing improved sampling over the standard approach. 

Diffusion maps can also be used to a compute the committor function~\cite{Lai2018}, which provides dynamical information about the connection between two metastable states and can be used as a reaction coordinate. Markov state models (MSM) can in principle be used to compute committor probabilities~\cite{prinz2011efficient}, but high dimensionality makes grid-based methods intractable. Similar work in this direction was done by~\cite{Lai2018,thiede2018galerkin,khoo2018solving}.
Diffusion-maps, especially the TMDmap~\cite{Banisch2018}, can be used for committor computations in high dimensions. The low computational complexity aids in the analysis of molecular trajectories and helps to unravel the dynamical behaviour at various temperatures. 

As a future work, the quality of the diffusion map approximation could be improved by introducing more sophisticated kernels or point-cloud approximations similarly to~\cite{Lai2018}. Also, diffusion maps could be extended to the approximation of generators of the underdamped Langevin dynamics.

\subsection{Extracting dynamical information from trajectory data}
\label{sec:34}

\noindent Once good CVs or metastable states have been identified, these can be used to extract dynamical information. Let us describe in this section the approach followed by Thiede {\it et al.} \cite{thiede2018galerkin}, which is based on a Galerkin projection of the infinitesimal generator.

The approach in \cite{thiede2018galerkin} builds on the MSM and related frameworks \cite{noe2013variational, williams2015data,schutte1999direct,molgedey1994separation, takano1995relaxation,hirao1997molecular,swope2004describing,prinz2011markov,giannakis2015spatiotemporal}. Dynamical statistics of interest are cast as solutions to equations involving the generator, i.e., the operator that describes the evolution of functions of the dynamics over infinitesimal times. Although the full generator cannot be determined in general, the equations can be solved by a Galerkin approximation.  In this approximation, the dynamical statistic of interest is expanded in terms of a basis, and its generator equation is reduced to a linear form. The contributing matrix elements (inner products of basis elements and the generator) can be estimated from short MD trajectories. A key challenge is to generate basis sets consistent with the boundary conditions. Thiede {\it et al.} \cite{thiede2018galerkin} considered two basis sets: indicator functions that reprise MSMs and diffusion maps \cite{Coifman2006}. The latter showed promise for capturing smoothly varying dynamical statistics, such as committors and mean first-passage times with fewer basis functions, but the efficiency of a given basis is likely to be problem specific.  Because the dynamical Galerkin approximation framework generalizes the notion of transition between states, the sampled configurations can be replaced by short trajectory segments. This allows treating memory that arises from incomplete description of the system by delay embedding \cite{takens1981detecting, aeyels1981generic}. This is an appealing alternative to extending the lag time in an MSM because it does not sacrifice time resolution. Going forward, it will be interesting to investigate whether variational methods akin to those for elucidating time scales \cite{noe2013variational,Mardt2018} can be developed to permit representation of the dynamical statistics in terms of nonlinear functions.

\subsection{Tackling both Markovian and non-Markovian cases: Free energy, friction and mass profiles extracted from short MD trajectories using Langevin models}
\label{sec:35}

\noindent In principle, the high-dimensional dynamics of a system composed by many atoms, when projected onto one (or a few) CV, can be modeled by a generalized Langevin equation~\cite{Zwanzig01,Luczka05}. Such stochastic differential equations contain several ingredients: a mass, a drift term corresponding to the mean force (gradient of the free energy landscape), a friction and a noise. Projecting on a low-dimensional space yields, in general, non-Markovian dynamics, except in the presence of time scale separation between CVs and bath coordinates and at coarse time resolution~\cite{Zwanzig01}.

Clearly, the construction of optimal Langevin models along meaningful reaction coordinates is appealing from several viewpoints \cite{camilloni2018}. On one side, the complex many-body dynamics is approximated by an equation that preserves physical intuition and is cheap to integrate. On the other side, exact kinetic rates - free from transition state theory approximations - between metastable states can be accessed more easily, by exploiting brute-force Langevin simulations or more elaborate methods~\cite{Hanggi90}. Compared to Markov state models, Langevin models are not restricted to Markovian dynamics and do not require the discretization of configuration space and the choice of a lag time, which are customary sources of errors.

For all these reasons, several algorithms have been developed to recast MD data into low-dimensional Langevin models~\cite{Straub87,Hummer03,Lange06,Hummer05,Horenko07,Micheletti08,Darve09,legoll2010effective,Schaudinnus15,Meloni16,Lesnicki16,Daldrop18}. Usually, with these techniques, the terms of the Langevin equation are estimated employing very long equilibrium MD trajectories that ergodically sample the whole relevant free energy landscape. Of course such data are seldom available in complex applications featuring rare events, strongly limiting the scope to the case of barriers smaller than a few $k_B T$. Tackling the more general case of limited sampling and non-equilibrium MD trajectories is much more involved~\cite{Zhang11}. 

A possible and simple solution to this challenge - especially in the context of rare events - has been proposed in Ref.~\citenum{perez2018}: the parameters of a generalized Langevin equation are optimized by minimizing the error between MD and Langevin  probability distributions $P(x,\dot x, t)$ along the reaction coordinate $x$. Such out-of-equilibrium distributions are estimated from a set of short unbiased trajectories initiated close to a barrier top (with random thermal velocities) and allowed to relax into the adjacent free energy minima, in the spirit of committor analysis (a preliminary exploration of putative transition state structures can be nowadays performed at a moderate cost using, e.g.,  the prejudice-free techniques of Ref.~\citenum{samanta2014,pietrucci2015,pipolo2017}).

Employing both benchmark models and solvated proline dipeptide as a test case, numerical evidence indicates that $\sim$100 short trajectories (of few picoseconds in the typical case of a small solute in water) encode all the information needed to reconstruct free energy, friction, and mass profiles \cite{perez2018}.
This approach, suitable also for high barriers of tens of $k_B T$ and non-Markovian dynamics, provides the thermodynamics and kinetics of activated processes in a conceptually direct way, employing only standard unbiased MD, at a competitive cost with respect to existing enhanced sampling methods.
Furthermore, the systematic construction of Langevin models for different choices of CVs starting from the same initial data could help in reaction coordinate optimization. 

\section{Application of machine learning techniques in biological systems and drug discovery}
\label{S:Applications}

\noindent Two of biology's biggest challenges are the prediction of protein structure based on its amino acid sequence, i.e., protein folding, as well as the dynamical conformational changes of the three-dimensional structure of proteins, i.e., protein dynamics. Beyond the actual problem of protein folding, which was recently set at a different basis after the breakthrough from AlphaFold and the impressive one million time faster Artificial Intelligence (AI) solution by AlQuraishi~\cite{AlQuraishi2019}, the prediction of protein dynamics and mechanism of action is possible through the use of MD simulations.

Recent advances in computer hardware and algorithms have led to simulations of protein dynamics of size and time lengths that are intrinsic to biological processes. Dynamics of protein plasticity and drug binding/unbinding mechanisms are a few of the key processes that we would ideally like to capture through these large scale simulations. However, the analysis and interpretation of the large amount of data that are produced by these simulations is complex and should be carefully considered \cite{Shaw2010}. 

As discussed in Section \ref{sec:32}, despite the ever-growing time and length scales of simulations, unbiased MD is not able to explore the whole kinetic landscape of complex systems and carefully chosen, meaningful CVs can be used to represent the free energy surface of these systems in order to reveal the regions of low energy, i.e., stable and metastable states, as well as the barriers, i.e., transition states, between these regions~\cite{Lange06,Schaudinnus15,Krivov2008}. ML approaches have recently started being used for the discovery of meaningful CVs \cite{chen2018collective,ribeiro2018reweighted,Mardt2018,Sultan2018,Brandt2018}, while iterative schemes where CVs are being updated based on new simulation data provide promising results for challenging systems \cite{Sultan2018,Chen2018,Trapl2019}. 

In this section, we first present an example of dimensionality reduction for building a Markov State Model for the study of lysine methyltransferase SETD8 (see Section~\ref{sec:CVs-MSM}). We next present some biological examples were adaptive MD/ML techniques can help gain access to non-crystallographic conformational states of disease-related proteins for drug discovery purposes (see Section~\ref{sec:ML-MD}). In Section~\ref{sec:Conf-Targeting}, we discuss the possibility of conformational-specific targeting of proteins using their metastable states as target conformations, while in Section~\ref{sec:Compound-effect} we give some examples were ML techniques applied in MD simulations can provide information about potential allosteric binding sites or protein activation mechanisms upon ligand binding.

\subsection{Selection of efficient collective variables for MSMs: the example of SETD8}
\label{sec:CVs-MSM}

\noindent Conformational changes in proteins span from thermal fluctuations of side chains and motions of active loops to major rearrangement of sub-domains, including unfolding and refolding processes \cite{Henzler2007}. The ability to unveil the mechanisms underlying protein function requires quantifying the importance of these motions for the process of interest or, in other words, obtaining a representative ensemble of conformations.

Besides the relevance for devising enhanced sampling strategies, the discovery of CVs is decisive when analyzing simulation data sets by using, for instance, Markov State Models. In this context, the conformational study of the protein methyltransferase SETD8, an epigenetic enzyme essential in the regulation of the cell cycle, was discussed in~\cite{Chen2018}.

SETD8 is characterized by a dynamically rich behavior, which has proven to be essential in enzymatic catalysis \cite{Schramm2011}. In~\cite{Chen2018} the authors combined experiments and simulation in an attempt to span the up-to-that-time unexplored configurational space of SETD8. Several new X-ray structures were obtained by trapping conformations with small-molecule ligands \cite{Lee2009}. These, in turn, were used to build hypothetical structures by manually combining fragments observed in experiments.

The set of initial configurations was used to seed independent MD simulations in explicit solvent, resulting in an extensive simulation database. The search of reaction coordinates was done in different spaces of residue-residue distances, logistic distances, and backbone dihedrals. These CVs, usually referred to as ``features'' in the MSMs literature, are arbitrary choices, that have been traditionally based on human intuition and heuristics \cite{Husic2016}. This is arguably the ``achilles heel'' of MSMs and has prompted the development of ML approaches to bypass human intervention \cite{Mardt2018,Wehmeyer2018}.

Although a set of features is already a space with much fewer dimensions than the full atomic coordinates, it is still a high dimensional system that cannot be handled with MSMs. This requires further dimensionality reduction, which can be done using, for instance, the time-lagged independent component analysis (tICA), discussed in Section~\ref{sec:slow_CV}. CVs obtained by tICA are linear combinations of features that, in principle, encompass the variance of the data while providing time scale separation. These are attributes of meaningful CVs \cite{Brandt2018}, which explains the consensus regarding tICA as a suitable strategy for building MSMs \cite{perez2013identification, Schwantes2013, Husic2016, Noe2017}. The stage regarding data representation ends with clustering the conformational snapshots into discrete states using unsupervised ML protocols, such as the k-centers and k-means methods \cite{Bowman_2014}. 

\begin{figure*}[tb]
\centering
\includegraphics[width=0.8\linewidth]{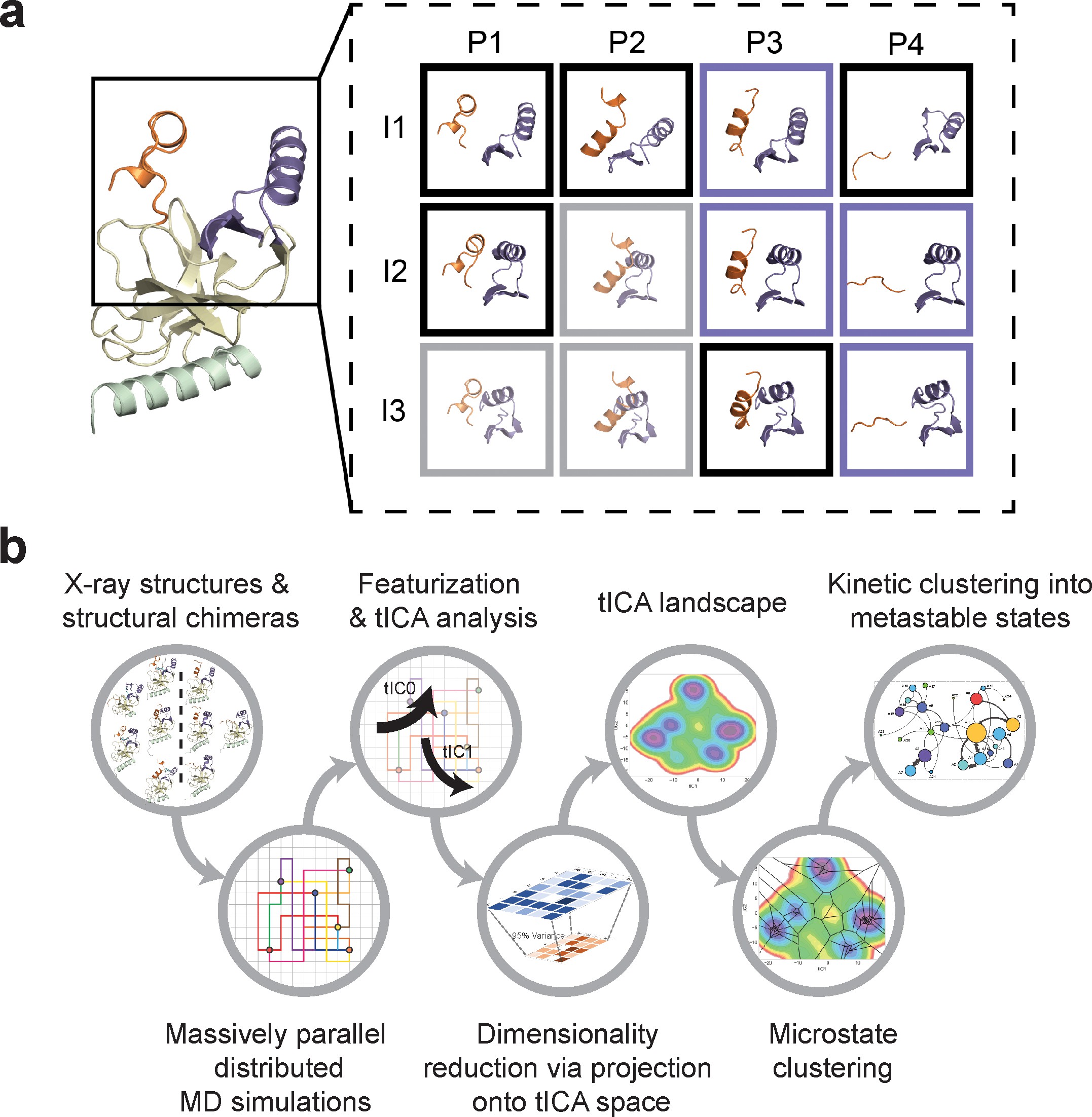}
\caption{Construction of conformational landscapes of apo- and SAM-bound SETD8 through diversely seeded, parallel molecular dynamics simulations and Markov state models.(a) Combinatorial construction of structural chimeras using crystallographically-derived conformations. (b) Workflow for dynamic conformational landscapes construction using MSM. For more information we refer the reader to the original publication \citenum{Chen2018}. (Image source: Ref.~\citenum{Chen2018}. Use permitted under the Creative Commons Attribution License CC BY 4.0., https://creativecommons.org/licenses/by/4.0/).}
\label{fig:SETD8}
\end{figure*}

Given the multiple subjective decisions involved in selecting features and algorithms to represent the database, MSMs building must be allied with validation strategies. In this context, Husic \textit{et al.} \cite{Husic2016} emphasize the importance of using a kinetically-motivated dimensionality reduction and cross-validation strategies to avoid over fitting. The study of SETD8 \cite{Chen2018} uses both structural and kinetic criteria, and 50:50 shuffle-split cross-validation scheme with random divisions of the data into training and test sets (see Figure~\ref{fig:SETD8}). As a result of such an extensive validation,  the specific study successfully quantified an ensemble of kinetically relevant macrostates which, in addition, were validated with experiments.

\subsection{Machine learning-driven MD simulations in drug discovery}
\label{sec:ML-MD}

\noindent The discovery of a new drug is a long, multi-step and expensive process. Any tool that can speed up any of the steps involved would have big implications down the entire drug discovery chain. Artificial intelligence is expected to significantly shape the future of many aspects of drug discovery during the forthcoming decades. It is already used to design evidence-based treatment plans for cancer patients, instantly analyze results from medical tests to escalate to the appropriate specialist immediately, and most recently to conduct scientific research for early-stage drug discovery.

Proteins, the most common drug targets, are dynamic molecular machineries whose function is intimately linked to their conformations. Destabilization of the subtle equilibrium of protein conformations can lead to severe pathologies, like in the well-known cases of KRAS G12X oncogenic mutations and prion disease. In this context, knowledge of the conformational landscape of targeted proteins would provide an outstanding advantage for the design of novel and original compounds stabilizing specific conformations of the protein~\cite{Wodak2019}. 

Experimentally, the protein conformational space is often limited to few conformations that have been prone to crystallize. The use of GPUs and massive computational resources has enabled for the \textit{in silico} alternative, MD simulations, to gain an important place in the first steps of drug discovery. Nevertheless, MD is limited to a few hundreds of microseconds of simulation, which limits the conformational space exploration.

New molecular modeling approaches combining MD simulations and ML techniques can help gain access to these non-crystallographic conformational states of a target protein. This knowledge would allow focusing on specific conformations of the protein in order to alter or restore its function. ML techniques can enable us to identify patterns in simulation data, build models that explain the different conformational states of a target and predict potential target-specific solutions for their druggability~\cite{Fiorin2013,Sultan2018,chen2018molecular,Brandt2018,ribeiro2018reweighted,Ung2018,Degiacomi2019,Diaz2019,Trapl2019}.

As discussed in Section \ref{sec:31}, good CVs can guide enhanced sampling MD simulations in order to gain insights into long timescale dynamics of biomolecular systems. The difficulty of the identification of such CVs and in most cases the complexity of their definition has limited the number of available software for this purpose. PLUMED is an open-source, community-developed library that has been widely used in enhanced-sampling simulations of complex biological systems in combination with many MD engines, e.g., Amber, GROMACS, NAMD, and OpenMM~\cite{Bonomi2009,Case2005,Berendsen1995,Phillips2005,Eastman2017}. Most importantly, PLUMED can be interfaced with the host code using an API, accessible from multiple languages, including C++ and Python). This last functionality is important for adaptive protocols used for the identification of optimal CVs using iterative learning algorithms based on well developed ML libraries like Keras~\cite{Chollet2015keras}, TensorFlow~\cite{tensorflow2015-whitepaper}, PyTorch~\cite{Paszke2017automatic} and Fastai~\cite{Howard2018fastai}. The MSM Builder package provides the user with software tools for predictive modeling of long timescale dynamics of biomolecular systems using statistical modeling to analyze physical simulations \cite{Harrigan2017}. Other tools that can be employed in MD/ML studies include among others MDTraj~\cite{McGibbon2015MDTraj}, ColVar module for VMD~\cite{Fiorin2013}, OpenPathSampling~\cite{ops2}.

\subsubsection{Conformational-specific targeting of proteins using cryptic binding sites}
\label{sec:Conf-Targeting}

\noindent Drugs are traditionally designed to bind to the primary active site of their biological targets in order to induce a therapeutic effect. However, the high similarity between the orthosteric pockets among most of the protein families, leads in several cases to adverse effects. A new emerging direction in drug discovery is the use of alternative, transient, non-orthosteric binding sites that are not apparent in the protein's known crystallographic conformations and where small molecules can bind and modulate the biological target’s function. 

By binding to non-orthosteric sites of proteins, allosteric inhibitors can also exhibit a better selectivity vs proteins from the same family, as illustrated by SAR156497, a highly selective inhibitor of Aurora kinases~\cite{Carry2016}. Well known drugs on the market work through this kind of mechanism of action (e.g., Lapatinib or Imatinib), but this mechanism was described \textit{a posteriori}. Moreover, there are approved allosteric modulator drugs such as Cinacalcet for the treatment of hyperparathyroidism and Maraviroc for the treatment of AIDS, as well as many candidates at different stages of development~\cite{drugbank,clinicaltrials}. Another aspect in targeting non-orthosteric pockets in drug discovery relies on the fact that allosteric inhibitors will not compete with endogenous ligands for binding, which can be critical when such endogenous ligands have very strong affinity for their protein. 

One of the successful efforts in this direction is the example of PI3K$\alpha$, where a novel non-orthosteric pocket was identified using molecular dynamics (MD) simulations \cite{Gkeka2014, Gkeka2015}. In \cite{Gkeka2014}, the authors used Functional Mode Analysis~\cite{Hub2009} and identified two dominant motions of PI3K$\alpha$ that influence both the active and allosteric pockets and are distinct between the wild-type protein and its oncogenic counterpart. Current work aims at extending this approach to other protein targets, where neural networks are employed in order to establish the link between oncogenic mutations and the protein's mode of action, with an ultimate goal to identify druggable mutant-specific conformations. 

Beyond single protein conformations, multimeric protein assembly also appears as a challenging area where ML could play a role in drug discovery. The recent example on TNF$\alpha$ for instance shows the importance of how subtle changes in protein conformation can translate into a distorded trimeric assembly of TNF$\alpha$, impacting downstream signaling of \MakeUppercase{TNFR1}. Small compounds stabilizing this asymmetrical TNF$\alpha$ trimer can then be designed to treat or prevent TNF$\alpha$-related diseases~\cite{TNF_patent_2016}.

\subsubsection{Compound-specific effect of binding}
\label{sec:Compound-effect}

\noindent Another promising direction in the drug discovery process is the compound-specific effect of protein binding~\cite{Barati2018,Feinberg2018}. For example, a small organic compound can be used to boost the enzymatic activity of a protein enzyme or evaluate allosteric binders by the stabilization of its active conformation. In finding allosteric binding sites, ML algorithms such as k-means and Markov Models can significantly help in reducing the dimensions of drug binding events. The connections between statistical mechanics principles, such as Boltzmann Machines, and the discovery of the binding sites in proteins can be insightful. As an example, one can run thousands of small trajectories of drug binding and unbinding events and learn the reaction coordinates using tICA (time-independent Component Analysis) in order to find the possible allosteric binding sites \cite{Barati2018}. These trajectories can be generated using different initial seeds (both different locations and orientations) and may range from 50 ns to 500 ns.
 
In the activation pathway of many proteins such as G Protein Coupled Receptors (GPCRs), the conformational changes are subtle and are limited to the sequential motion of residue switches triggering a signal from ligand to intracellular motifs. Finding these intricate motions in high dimensional space requires ML techniques to reduce the system's dimensions \cite{Feinberg2018}. Among these methods, variational autoencoders (VAE) and tICA (sparse or kernel) can be used to achieve learning and finding the reaction coordinates for such complex proteins.

\section{Concluding remarks and perspective}
\label{S:Perspective}

\noindent Let us conclude this review by presenting some global perspectives on the interactions between machine learning approaches and molecular simulation, which are common to all the situations we discussed -- from devising numerical potentials based on ab-initio reference data to the identification of collective variables in actual simulation of biological proteins.

First, we have seen that the aims of the coarse-graining procedures may be very different in nature. From the material presented in this review, one can identify three major purposes: (1) {\em a modeling objective}: using machine learning techniques to improve models, for instance by better representing force fields and potential energy surfaces; (2) {\em a numerical objective}: improving the efficiency of numerical methods, for instance by devising good collective variables to be used in conjunction with enhanced sampling techniques, such as free energy biased sampling techniques; (3) {\em a data analysis objective}: providing an efficient post-processing tool, as for instance a Markov state model to interpret the raw simulation data from molecular dynamics and identify states of interest.

Concerning the choice of the learning methods, some common trends are shared by all methods, namely ensuring that one has access to a sufficiently rich database (sufficient variability of configurations for force fields, long reactive trajectories to identify CVs) and representing correctly the data (starting possibly with some putative CVs/descriptors, and then using some regression from there to sparsify/optimally combine these initial guesses). The precise choice of the learning method and the reduced model to work with, however, depend very much on the goal and priority of the user, and the system under consideration. The priority can be {\em the accuracy} (being as precise and as close as possible to some reference model, e.g., all-atom results when coarse-graining, or reproducing DFT energies when constructing numerical potentials), {\em the transferability} (learning how to coarse-grain small systems and extending the method to larger ones, learning energies at a given temperature and using the potential at another one) or the CPU/GPU {\em computational cost}. 

When using black box learning techniques, based for example on neural networks, a problem which is often raised is the {\em interpretability} of the result. This is discussed for example in~\cite{jung2019artificial} which attempts to reconcile machine learning models (specifically a neural network approach to optimal reaction coordinates) with physical insight by means of symbolic regression techniques, also known as genetic programming. Such techniques appear very promising for the future, being able to distill fundamental natural laws from numerical data \cite{schmidt2009distilling}.

Another important element is the {\em reproducibility} of the results: one should favor approaches which are easy enough to cross-check and to repeat on various architectures. This also requires the researchers to ensure that the coarse-graining technique they propose yield robust results. For example, the results should not depend on the initial weights in a neural network, or on the sampled point used as inputs. Finally, this includes considering well established databases, or making databases available to other users/developers; and also relying on standard and well maintained packages when using external libraries.

One idea which would help setting up common benchmarks and/or agreeing on common aims/priorities would be to organize some competition or prediction contest, which should ideally be simple enough so that even small groups can participate since this requires agreeing on common goals. Setting up the rules of such a competition would already be quite an achievement. Another important idea would be to emphasize transferability in all approaches, and more systematically work with some databases of some sort and then test on different databases.


\begin{acknowledgements}
This review paper was written following a CECAM (Centre Européen de Calcul Atomique et Moléculaire) discussion meeting, hosted at the Sanofi Campus of Gentilly. The authors thank the CECAM as well as Sanofi for making this event possible. Moreover, the PG, GS and TL thank Dr. Marc Bianciotto for his proof reading and feedback. 
\end{acknowledgements}


\section{Bibliography}
\bibliography{Article}

\onecolumn


\end{document}